\documentclass[amssymb,prb,amsmath,superscriptaddress,twocolumn,showpacs,aps]{revtex4}
\usepackage{bm}
\usepackage{float}
\usepackage{graphics}
\usepackage{graphicx}
\usepackage{amsfonts}
\usepackage{amsmath}
\usepackage{amssymb}
\usepackage{graphicx}
\usepackage[stable]{footmisc}
\usepackage[english]{babel}
\begin{document}

\preprint{HEP/123-qed}
\title[Short title for running header]{Measuring roughness of buried interfaces by sputter depth profiling}
\author{S.V. Baryshev}
\email{sbaryshev@anl.gov} \email{sergey.v.baryshev@gmail.com}
\affiliation{Material Science Division, Argonne National
Laboratory, 9700 S. Cass Ave., Argonne, IL 60439}
\author{J.A. Klug}
\affiliation{Material Science Division, Argonne National
Laboratory, 9700 S. Cass Ave., Argonne, IL 60439}
\author{A.V. Zinovev}
\affiliation{Material Science Division, Argonne National
Laboratory, 9700 S. Cass Ave., Argonne, IL 60439}
\author{C.E. Tripa}
\affiliation{Material Science Division, Argonne National
Laboratory, 9700 S. Cass Ave., Argonne, IL 60439}
\author{J.W. Elam}
\affiliation{Energy Systems Division, Argonne National Laboratory,
9700 S. Cass Ave., Argonne, IL 60439}
\author{I.V. Veryovkin}
\email{verigo@anl.gov}
\affiliation{Material Science Division,
Argonne National Laboratory, 9700 S. Cass Ave., Argonne, IL 60439}


\begin{abstract}
In this communication, we report results of a high resolution
sputter depth profiling analysis of a stack of 16 alternating MgO
and ZnO nanolayers grown by atomic layer deposition (ALD) with
thickness of $\sim$5.5 nm per layer. We used an improved dual beam
approach featuring a low energy normally incident direct current
sputtering/milling ion beam (first beam). Intensities of
$^{24}$Mg$^+$ and $^{64}$Zn$^+$ secondary ions generated by a
pulsed analysis ion beam (second beam) were measured as a function
of sample depth by time-of-flight secondary ion mass spectrometry
(TOF SIMS).

\par Experimental results of this dual beam TOF SIMS depth profiling
processed in the framework of the mixing-roughness-information
(MRI) model formalism demonstrate that such an approach is capable
of providing structural information for layers just a few nm
thick. Namely, it was established that the interfacial roughness
of the MgO/ZnO multilayer structure equals 1.5 nm. This finding by
TOF SIMS was cross-validated by independent measurements with
specular X-ray reflectivity (XRR) technique. In addition, the TOF
SIMS-MRI analysis suggests that the obtained 1.5 nm roughness
should be attributed to the native roughness (jagged type) of the
interface rather than to interdiffusion at the interface during
the ALD synthesis.

\end{abstract}

\maketitle

\section{Introduction}

Quality of interfaces in multilayered materials is of great
interest and importance, since the way layers adhere to each other
greatly affects their performance: spin-based electronics,\cite{1}
giant magnetoresistance structures/devices,\cite{2} high-$k$
multilayer materials,\cite{3} artificial Bragg crystals,\cite{4}
heat transfer in nanostructures,\cite{5} and others.

\par A powerful and efficient set of methods commonly used to
determine (i) single layer thickness, (ii) roughness and (iii)
density in a multilayer system combines the X-ray reflectivity
(XRR) method with ellipsometry. XRR technique uses the values of
(i), (ii), and (iii) as fitting parameters jointly with some type
of boundary conditions to solve a general inverse problem of wave
scattering.\cite{6,7} XRR results can be partially validated by
optical ellipsometry, which is also based on solving a similar
inverse problem. It relies on refractive index (a function of
density) and a layer thickness in a single/multiple layered
structure, and depends on boundary conditions as well. This makes
the data fitting model sensitive and dependent on many free
parameters, which can grow fast when the number of layers in the
layered structure increases. Moreover, such indirect techniques
based on measuring optical responses of a material may become
inefficient if one needs to work with three-layer stacks (or even
more general case of $n$ layers) where repetition of certain
layers in an $n$-layer stack varies, as, for instance, occurs in
the case of metamaterials.\cite{8,9} A problem will also exist for
multilayer semiconductor heterostructures, where the composition
variation between nearest layers is about the same with a few
percent difference in solid solution content.\cite{10} Hence, the
change in density is so small that the sensitivity of the XRR
method may be insufficient for characterization of such materials.
Thus, the search for a more direct and established approach to
characterization of nm-thick layers in a variety of multilayered
structures, which is sensitive and independent on optical
properties, is highly desirable. It can complement XRR and
ellipsometry techniques, and in some cases might do the job, which
is beyond their capabilities.

\par Mass spectrometry retains its importance in basic
physics,\cite{11} chemistry,\cite{12} and in materials
science\cite{13} due to a number of unique advantages, such as
sensitivity to isotopic ratio, trace amounts of
elemental/molecular species on the surface and in the bulk, etc.
In this context, an extension of its capabilities is of great
interest. In the present paper, we demonstrate that properly
designed mass spectrometry of secondary ion species in sputter
depth profiling experiments can successfully reveal structural
features of nanolayered structures and characterize their surface
and interfaces. To this end, an improved dual beam secondary ion
mass spectrometry (SIMS) with low energy normal incidence
sputtering/milling was applied to depth profile a layered
MgO/ZnO$\times$8 structure grown by atomic layer deposition. The
obtained depth profiles were processed in terms of
mixing-roughness-information (MRI) model. This model has
analytical solutions with, in case of SIMS, two variable
parameters having straightforward physical interpretation: the ion
beam mixing length and the roughness. The SIMS-MRI results are
compared to structural parameters obtained by specular XRR
measurements to confirm the correctness of made conclusions.

\section{Samples and Experimental}

\par A layered structure $\mid$5.5-nm MgO/5.5-nm ZnO$\mid$$\times$8 was grown by ALD
on a Si substrate using established precursor chemistries for
MgO\cite{14} and ZnO\cite{15,16} and characterized by X-ray
diffraction and ellipsometry. The structural data obtained from
these measurements indicated that MgO layers are amorphous and ZnO
layers are polycrystalline in the wurtzite phase. The roughness of
the Si substrate was $\sim$0.3 nm. The initial layer-to-layer
mixing due to thermal diffusion during growth at $T$=473 K is
expected to be extremely low. Specular X-ray reflectivity
measurements were made with a Philips X'Pert Pro MRD
diffractometer using Cu K$\alpha$ radiation ($\lambda$=1.5418
${\AA}$) and operated at 30 kV/40 mA. The incident X-ray beam was
conditioned by a 60 mm graded parabolic W/Si mirror with a
0.8$^\circ$ acceptance angle and a 1/32$^\circ$ divergence slit.
The reflected beam was collected with a PW3011/20 sealed
proportional point detector positioned behind a 0.27$^\circ$
parallel plate collimator.

\par Secondary ion mass spectrometry studies were performed in a
custom-designed SARISA (Surface Analysis by Resonance Ionization
of Sputtered Atoms) instrument in the Materials Science Division
at Argonne National Laboratory.\cite{17} SARISA combines two
independently optimized Ar$^+$ ion beams: one of low energy (a few
hundred eV) and normal incidence for ultimate depth resolution ion
milling/sputtering, and another for elemental time-of-flight (TOF)
SIMS analysis with high lateral resolution. This arrangement is
based on the known powerful dual beam approach to depth
profiling\cite{18} and can be dubbed
\emph{gentle}DB\footnotemark[0].\footnotetext[0]{\\\emph{gentle}
stands for low energy normal incidence sputtering, \emph{lenis},
which is \emph{gentle/soft} in Latin} There are three main
advantages of this setup. First, depth resolution is controlled by
the milling beam if the parameter $\alpha$ is much smaller than
unity, or $\alpha\ll 1$ (see below). Second and third, normal
incidence permits varying the impact energy of primary ions by
target bias and does not introduce additional roughening of a
sample surface.\cite{19,20}

\begin{figure}[t] \centering
\includegraphics[width=6.cm]{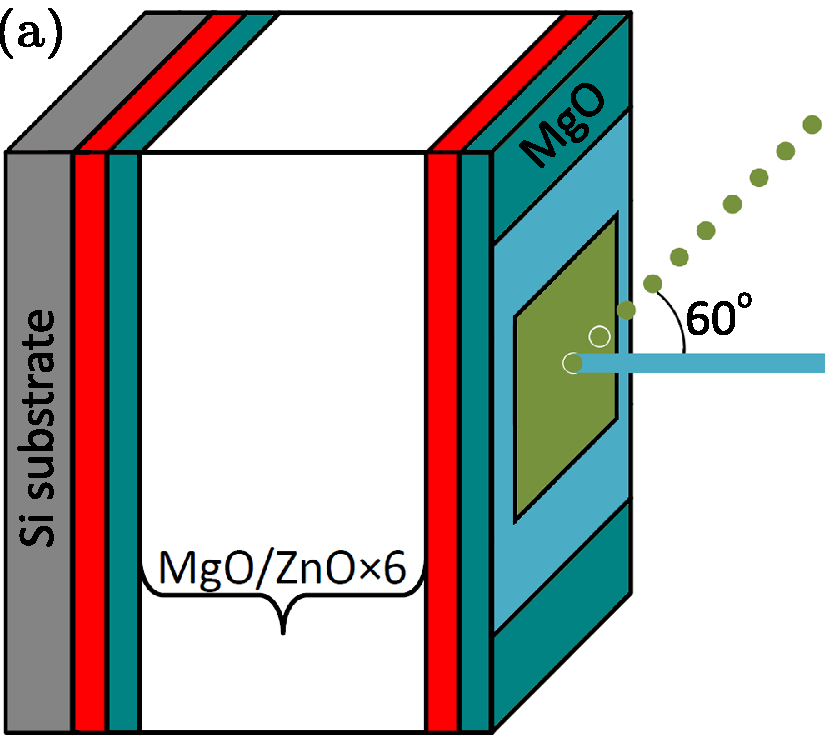}

\

\includegraphics[width=7.cm]{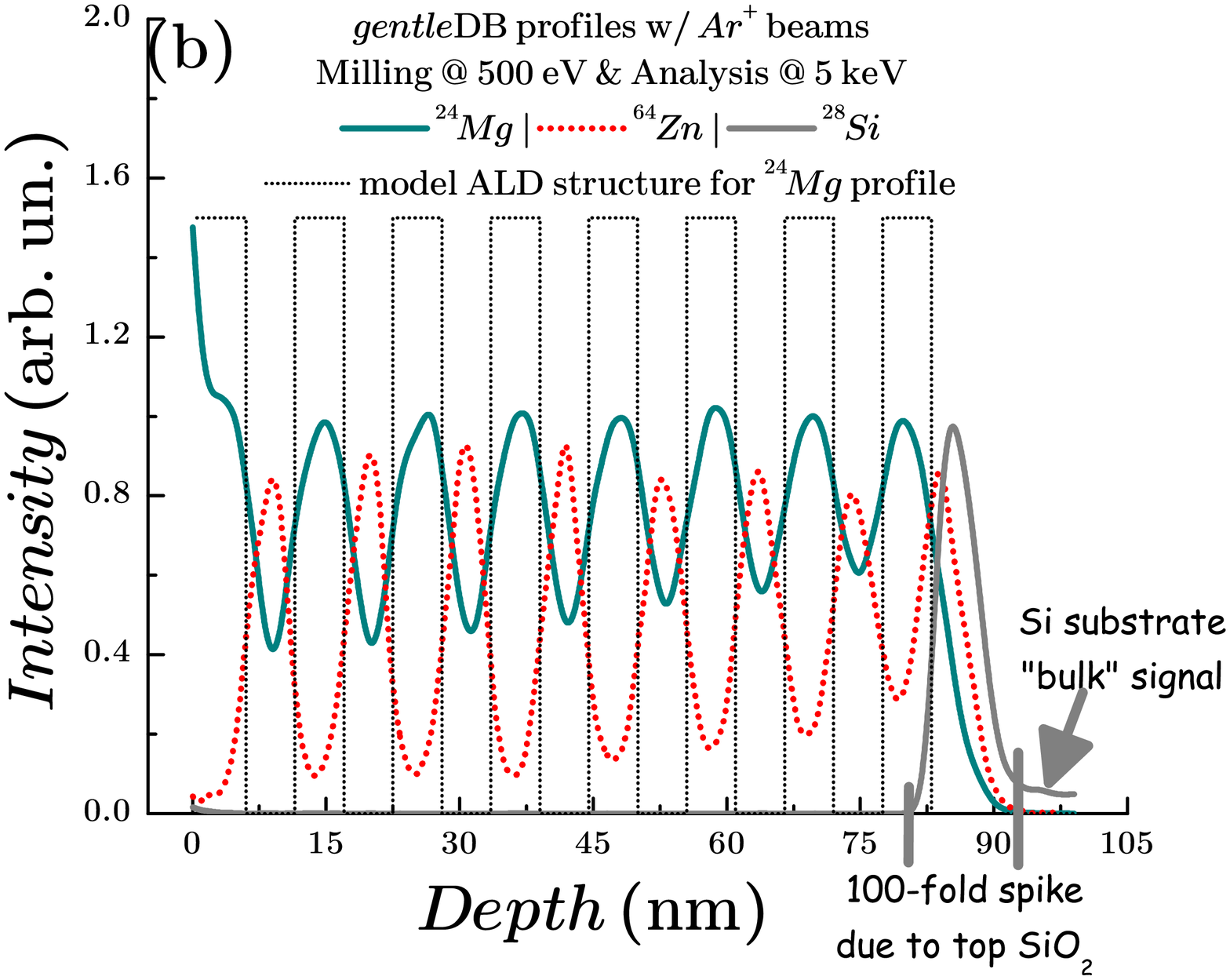}
\caption{(a) Simplified experimental setup: 8 MgO (cyan)/ZnO (red)
pairs, superimposed and centered blue and green squares are
craters from a raster scanned direct current normal incidence
milling Ar$^+$ beam (impinging solid blue line), and a raster
scanned area of a pulsed probing Ar$^+$ beam (impinging green
dotted line), respectively. (b) A full \emph{gentle}DB SIMS depth
profiles of the MgO/ZnO$\times$8 ALD structure on a Si substrate
obtained at the 500 eV milling Ar$^+$ beam combined with the 5 keV
probing Ar$^+$ beam.}
\end{figure}

\par Elemental depth profiles, concentration/intensity versus depth,
of the sandwich structure were obtained by a sequence of
alternating cycles. Ion milling by a raster scanned primary direct
current Ar$^+$ ion beam at 500 eV and normal incidence\cite{17} is
followed by TOF SIMS analysis of revealed subsurface (various
depths, from surface down to substrate) by a raster scanned pulsed
(200 ns long) Ar$^+$ ion beam at 5 keV energy and 60$^\circ$
incidence with respect to the target normal, as shown
schematically in Fig.1a. This pair of cycles is repeated multiple
times until the Si substrate is reached, which is monitored by
Si$^+$ peak intensity in a mass spectrum. In the depth profile,
the Si substrate grows in as a spike attributed to $\sim$100-fold
enhanced secondary ion yield due to presence of SiO$_2$. It then
stabilizes at a lower constant level, as shown in Fig.1b. The
depth resolution is controlled ultimately by the milling beam
characteristics. This condition is fulfilled if the parameter
called effective erosion efficiency $E=Y\cdot j\cdot t/e$ specific
for the milling beam is much higher than that of the analytical
beam, or $\alpha=E_a/E_s\ll1$. Here, $Y$ is a sputtering yield,
which depends on a beam energy $\varepsilon$, the primary
projectile species and an angle they attack the surface, $j$ is
the current density, $t$ is the total sputtering time during which
an ion beam is on. In our experimental setup $\alpha$ was
$\sim$10$^{-5}$.

\par The milling beam was digitally raster scanned over a square
area of $\sim$1 mm$^2$ (blue square in Fig.1a), while the
analytical beam was raster scanned over a square of
$\sim$500$\times$500 $\mu$m$^2$ (green square in Fig.1a). Both
raster areas were precisely overlapped by using an \emph{in situ}
Schwarzschild microscope,\cite{17} and \emph{ex situ} by white
light profilometry.\cite{21}

\par Ion beam currents were measured and focused \emph{in situ} by a custom
made graphite Faraday cup (FC) consisting of an internal pin and
inlet holes of 250 $\mu$m dia. on the mask. Such a FC design
permits to control focusing conditions of an ion beam. Ion beam
profiles were checked and known to have a symmetric Gaussian
distribution by burning dents in soft materials and profiling the
dents by \emph{ex situ} white light profilometry.\cite{21} It
allows one to precisely calculate a current density of an ion beam
using a known dc current value measured on the FC. The milling
beam parameters at 500 eV were 1 $\mu$A dc and a FWHM of
Gaussian-like beam profile of $\sim$150 $\mu$m. The analytical
beam characteristics at 5 keV were 300 nA dc and a FWHM of
Gaussian-like beam profile of $\sim$40 $\mu$m, and then reduced
down to 30nA by choosing an appropriate aperture.

\par The measured energy spread $\Delta\varepsilon$ of our low energy column\cite{17} is 23
eV yielding $\Delta\varepsilon/\varepsilon\sim 5\times 10^{-2}$ at
500 eV milling energy. Thus, the milling beam can be considered as
monoenergetic with high precision.

\section{Results, Discussion and Conclusions}

\par Fig.1b demonstrates SIMS depth profiles of Mg$^+$ and Zn$^+$ obtained
by \emph{gentle}DB at 500 eV ion milling, reflecting the full set
of peaks in the periodical structure, 8 peaks for MgO and ZnO
each. Fig.2a is a high-resolution profile obtained under the same
conditions as the profiles in Fig.1b, but at small ion milling
increments, so that the depth difference between two consecutive
points is $\sim$0.2 nm, corresponding to monolayer thickness for
wurtzite ZnO.

\par We used the high-resolution Zn$^+$ depth profile for further
evaluation of structure features of the multilayer ALD stack. This
was accomplished in the terms of an analytical approach called
mixing-roughness-information (MRI) model.\cite{22,23} In essence,
the model has 3 physically meaningful parameters. Mixing is
characterized by the ion beam mixing length $w$, a scale at which
two perfect layers of different compositions with perfect abrupt
interface experience mutually uniform interpenetration under
nonreactive gas bombardment. Roughness is the root-mean-square
(rms) roughness $\sigma$. By the term "information", the depth of
collecting information $\lambda$ is meant, which is the same as
the escape depth commonly used in surface analysis techniques. In
general, $\lambda$ is a variable parameter, but in the case of
SIMS it can be fixed at a value of 1-2 monolayers.\cite{24} The
effects of the two main processes, mixing and roughness, on the
resulting peak depth profile are shown in Fig.2b. It is seen that,
in the case of an ideal rectangular layer, the ion beam induced
mixing leads to asymmetric depth profile, which may get additional
broadening and symmetrization due to the rms roughness, depending
on the interplay between $w$ and $\sigma$ values. The effect of
the information depth $\lambda$ is not incorporated, since it is
expected to be negligible.

\par The green solid curve in Fig.2a superimposed on the Zn$^+$
depth profile is the best fit of the experimental curve by the MRI
model. The fit parameters were as follows: $w$ is 0.4 nm and
$\sigma$ is 1.5 nm, while $\lambda$ was fixed at 0.2 nm,
corresponding to the wurtzite ZnO monolayer thickness. The flat
thickness of the layer $d$, which is a thickness of an ideal flat
layer (see Fig.2b) defining boundary conditions for the model,
appeared to be 3 nm yielding the total single ALD layer thickness
to be of 6 nm. The sharpness of the last interface between the Si
substrate and the ZnO layer can also be estimated. An up-slope of
the Si signal shown in Fig.1b results in a roughness of $\sim$0.4
nm.

\begin{figure}[t] \centering
\includegraphics[width=7.cm]{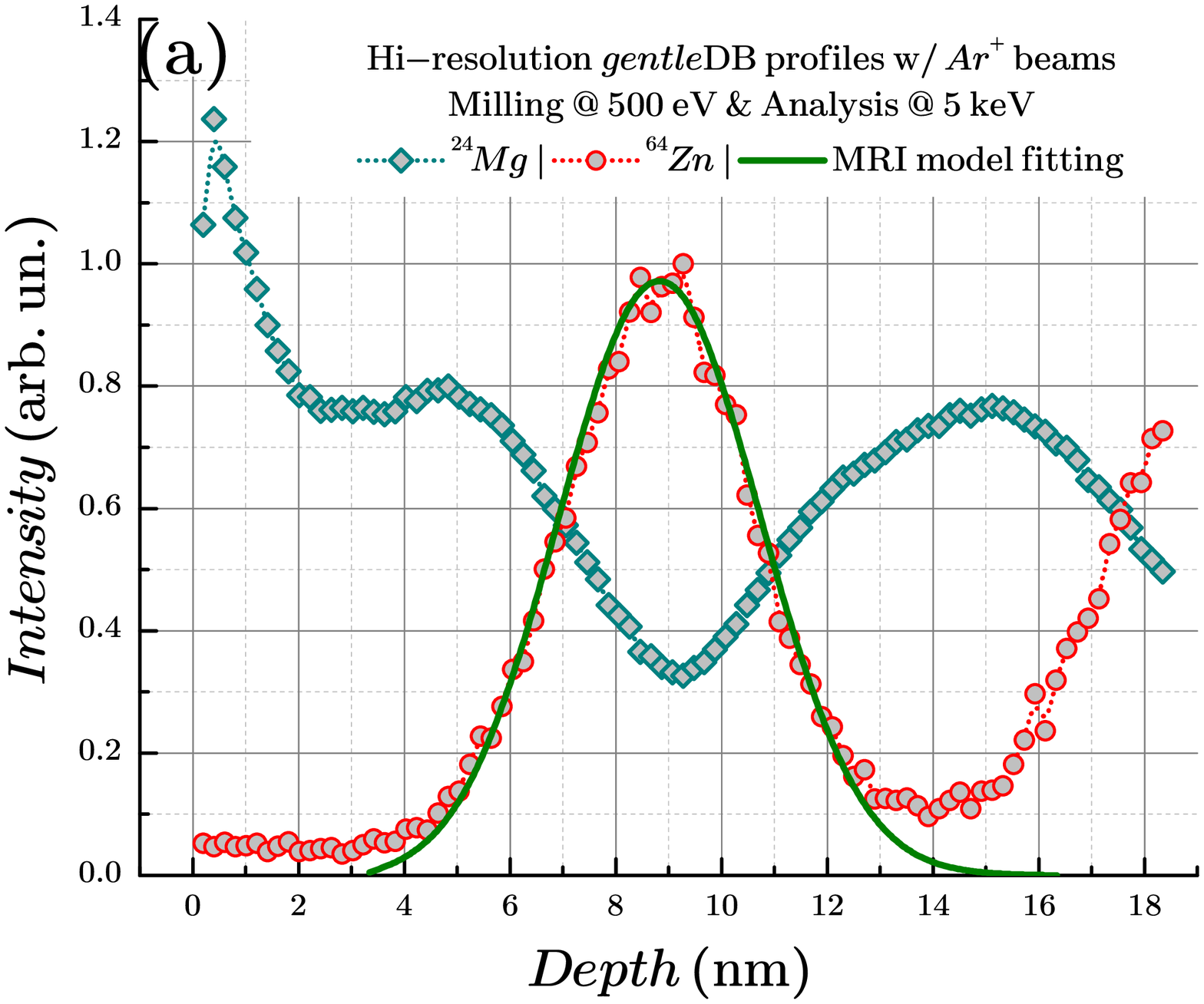}

\

\includegraphics[width=6.cm]{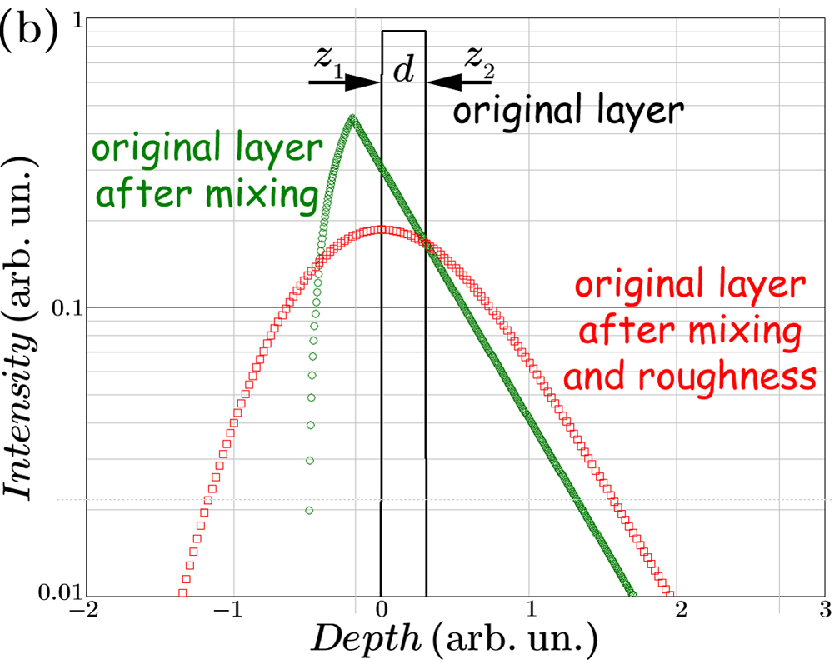}
\caption{(a) High-resolution \emph{gentle}DB SIMS depth profiles
of MgO/ZnO top layers at 500 eV milling Ar$^+$ beam and 5 keV
analytical Ar$^+$ beam. Mg signal is shown in cyan (open
diamonds), and Zn is shown in red (open circles). The green solid
curve is the best MRI model fit. (b) Principles of the MRI model
describing how different distortion sources affect an ideal,
rectangular (flat) layer profile of thickness $d$. The effect of
the escape depth parameter, negligible in the case of SIMS, would
slightly broaden the red curve (open squares) and is not shown.}
\end{figure}

\par To cross-validate the results obtained by MRI modeling of the
experimental data, we performed specular XRR measurements on a
MgO/ZnO sample prepared under identical conditions. The use of XRR
in this case seems to be a valid approach, since its sensitivity
to fluctuations of the thickness (that can range in characteristic
length scale) at the wavelength of $\sim$1.5 $\AA$ can be
anticipated to be similar to that of SIMS based on a single ion
penetration in the fine interfacial structure. Fig.3 represents a
typical specular XRR profile obtained for the MgO/ZnO$\times$8
sandwich. Fitting of experimental data was performed with a
commercial software package (Panalytical X'Pert Reflectivity)
which makes use of the Parratt recursion formalism for
reflectivity.\cite{25} Due to the relatively large number of
discreet layers present in the MgO/ZnO$\times$8 heterostructure,
several periodic constraints were applied to reduce the set of
fitting parameters which would otherwise be required. A simple
ideally periodic model assuming identical ZnO/MgO bilayers failed
to reasonably approximate the data. A second model where the first
ZnO layer and last MgO layer were treated discreetly as described
by Jensen \emph{et al}\cite{7} also failed to produce adequate
fits to the experimental data. However, simulations where the
density, $\rho$, thickness, and roughness, $\sigma$, of each
subsequent ZnO and MgO layer were constrained to increase linearly
with increasing distance from the substrate provided a good match
to the experimental reflectivity curve. The experimental fit was
further optimized by allowing each $\rho$, thickness, and $\sigma$
to vary from the linear model within reasonable physical
constraints (e.g. bulk density values). The resulting reflectivity
fits confirmed the parameters values extracted from SIMS data: the
average thicknesses of ZnO and MgO appeared to be 6.4 nm and 6.1
nm, with average roughness values of 1.1$\pm$0.2 nm and
1.6$\pm$0.3 nm, respectively. Parameters of the SiO$_2$ layer on
the top of the Si substrate were also of the expected values:
thickness of 1.2 nm, $\sigma$=0.3 nm.

\begin{figure}[t] \centering
\includegraphics[width=7.cm]{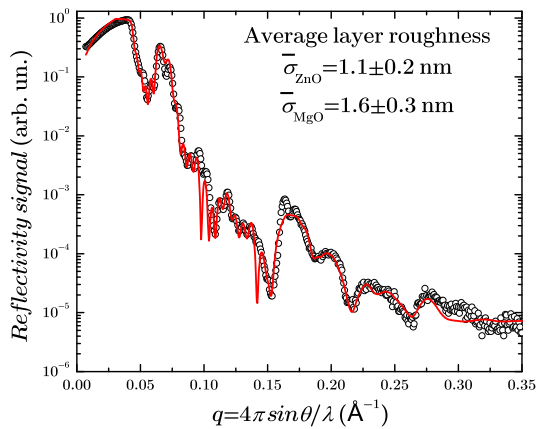}
\caption{A specular XRR profile
of the MgO/ZnO$\times$8 ALD structure on the Si substrate obtained
at 1.5418 $\AA$ irradiation. The black circles are the
experimental reflectivity curve, while the simulated reflectivity
curve is represented by the red solid line.}
\end{figure}

\par Thus, XRR data independently proves that SIMS analysis along
with the chemical information is capable of revealing detailed
internal interface structure, when distortions/artifacts related
to a profiling procedure are minimized to a scale, much smaller
than that of a feature to be investigated. In the case of
\emph{gentle}DB approach, this means that the extracted mixing
parameter $w$=0.4 nm is of the same order as the inherent escape
depth of a secondary ion, a physical limitation of the technique.
Therefore, the found 1.5 nm interfacial roughness of the MgO/ZnO
layers is due to an ultra-short ion beam mixing length/lattice
disturbance in \emph{gentle}DB profiling procedure. XRR also
proves that the obtained 1.5 nm roughness is reasonable and falls
well within the range provided by other experimental methods, both
in this study and other studies of layered laminate oxides
performed by AFM/STM,\cite{26} XRR\cite{3,7} and TEM.\cite{3}

\par Let us extend the finding by \emph{gentle}DB SIMS in conjunction with
MRI results. First of all, in the context of the advantages of the
low energy orthogonal sputtering described above, the rms
roughness $\sigma$ is believed to be a native interface roughness
of the grown ALD structure. In addition, it is known that XRR
cannot distinguish between the interfacial roughness and the
interdiffusion.\cite{7} One may assume that the interdiffusion at
the interface is a process similar in some way to the ion beam
mixing, which normally causes a mutually uniform layers
interpenetration,\cite{23} thus blurring the initially sharp
interfaces and forming a mixed-up interlayer. If the
interdiffusion is pronounced, the MRI model suggests that the
\emph{gentle}DB SIMS depth profiles (keeping in mind the limit of
$w$ going down to zero) would look more asymmetric, closer to the
green curve shown in Fig.2b. For this particular MgO/ZnO structure
grown at only 473 K, the shapes of the Mg$^+$ and Zn$^+$ peaks,
which are broad and nearly symmetric, in measured depth profiles
suggest that the estimated 1.5 nm roughness corresponds to native
roughness (i.e., "jagged" junction), rather than to
interdiffusion, or that the interdiffusion does not exceed 0.4 nm
from the total 1.5 nm.

\par To summarize, \emph{gentle}DB SIMS, in conjunction with the MRI model,
is a direct, but destructive and relatively slow method, which
does not require prior assumptions regarding the sample. On the
contrary, the specular XRR is indirect, but nondestructive and
relatively fast method, which is model dependent and needs prior
information about the sample. Thus, when both techniques are
applied to characterize buried interfaces in nanolayered
materials, they complement and enhance each other, and yield
comprehensive physical/structural and chemical information on the
material: interfacial roughness distinguishing interdiffusion from
native/jagged roughness, thickness, density, as well as elemental
depth profiles, including trace impurities, redistribution of
isotopes in isotopically modulated structures,\cite{27} and many
other possible applications.

\section{Acknowledgments}

\par This work was supported under Contract No.
DE-AC02-06CH11357 between UChicago Argonne, LLC and the U.S.
Department of Energy.

\end{document}